# Characterization and improvement of axial and radial stiffness of contactless thrust superconducting magnetic bearings

Ignacio Valiente-Blanco [2)] · Efren Diez-Jimenez*[1)] · Cristian Cristache[2)] · Marco A. Alvarez-Valenzuela[1)] and Jose L. Perez-Diaz[1)]

[1)] Department of Mechanical Engineering, Universidad Carlos III de Madrid
[2)] Instituto Pedro Juan de Lastanosa, Universidad Carlos III de Madrid
Avda. Universidad 30, 28911 Leganes, Spain.
*Corresponding author: ediez@ing.uc3m.es

**Abstract** Contactless bearings based in both permanent magnets and superconducting magnetic levitation are interesting to avoid all the tribological problems associated with contact at very low temperature. Superconducting magnetic bearings (SMBs) find application in mechanical engineering where lack of contact is a requirement or an advantage. In comparison with active magnetic bearings (AMBs), SMBs solve their inherent instability and require less complex control strategies and electronics. However, one of the current limitations is their low mechanical stiffness. Force densities are considerably lower in SMB than the state-of-art values for AMB. This fact certainly limits their applications. In this paper, we summarize some key strategies for the improvement of radial and axial stiffness among with a description of some SMB in terms of load capability or force relaxation and discuss their advantages and disadvantages. A set of rules for the mechanical design of high-stiffness thrust superconducting magnetic bearings are proposed and experimentally demonstrated

## 1. Introduction

Contactless bearings based in both permanent magnets and superconducting magnetic levitation are interesting to avoid all the tribological problems associated with contact at very low temperature [1],[2] such as friction, wear or energy losses [3], [4],[5],[6] . In addition, insofar there is no contact between the moving parts, lubrication is not required. This presents plenty of advantages from the maintenance, reliability and lifetime point of view [7],[8],[9].

Superconducting magnetic bearings (SMBs) find application in many fields where lack of contact is a requirement or an advantage such as in flywheel systems and transportation [10], cryogenic turbine flow meters, cryocoolers and sensitive gyroscopes for space applications [4] among other applications for cryogenic and space mechanisms [12–19]. Very recently, they have been proposed for solving tribological problems in dry-lubricated harmonic drives for space applications [20].

In comparison with active magnetic bearings (AMBs), SMBs solve their inherent instability and require less complex control strategies and electronics. However, one of the current limitations is their low mechanical stiffness. Force densities are considerably lower in SMB than the state-of-art values for AMB [21],[22]. This fact certainly limits their applications [23].

Despite that interesting ideas for radial and axial stiffness optimization of thrust SMB can be derive from previous works, further analysis is still required. In this paper, we summarize some key strategies for the improvement of radial and axial stiffness together with a description of some specific SMBs in terms of load capability or force relaxation and discuss their advantages and disadvantages.

## 2. Bearing design

The mechanical interaction between magnets and superconductor has been previously deeply analyzed [24–29]. There are some models that are useful to describe this interaction and that can be applied in finite elements programs as it is commonly used in mechanical engineering [30–32]. Although these models can be very useful for static loads calculations, they are still limited when dynamics are transient effects are taken into account. Hence the real experimentation of SMBs design is still necessary.

Among superconducting bearings, thrust bearings have been typically considered in order to bear high axial loads. For most configurations in the literature, a superconducting thrust bearing is composed by a ring permanent magnet (PM) levitating over a disk superconductor (SC).

In order to design a thrust bearing, load requirements should be considered in a first place. In flywheel systems, for instance, the load capability is greatly important because it will determine the maximum energy that can be stored [10]. The elastic nature of SMB means that higher load capabilities can be achieved for larger axial displacements of the PM levitating over the superconductor, as shown for example in Fig. 5. Therefore, larger values of the height of field cooling position (HFC), defined as the initial gap between the PM and the superconductor, tend to provide higher load capabilities.

Stiffness is a main parameter to take into account for certain applications. Despite that larger values of the HFC can provide greater load capabilities, lower axial and radial stiffness can be obtained for the same bearing if the HFC is increased (see Fig. 6). Stiffness are greatly relevant especially in applications where space is limited and stability is the most demanding requirement, as for example, in the MADGRIVE FP7 Project [20]. Previous works that analyze radial and axial forces in thrust SMB have reported typical values of radial stiffness of about a few N/mm [11], [33].

The influence of the shape, size and arrangement of PMs and SCs in a thrust bearing has been reported to be also very relevant for the performance of the bearing [34–37]. In general, multipole arrangements in a SMB have been previously explored showing good results in improving the load capability of SMB. The basic concept is that, by an appropriate configuration of the PMs in a SMB, higher flux densities and higher magnetic gradients can be applied to the SC, thus increasing its stiffness and load capability. However, further information of the mechanical behavior and performance of such bearings is required.

In a paper from Sotelo et al. [38], the potential of adapting a Halbach configuration of the PMs in improving the radial and axial stiffness of thrust bearings has been investigated. However, only experimental validation of the FEM calculated axial stiffness is given in this paper. Despite these efforts in analyzing multipole thrust bearings, a deeper analysis of the mechanical behavior of them is still required.

Finally, some other parameters as the force relaxation or the rotation losses must be considered from the point of view of the mechanical engineer for an optimum design of the SMB [39–41].

The foregoing concepts will be experimentally validated in the following section.

### 3. Experimental set up and procedure

SMBs are typically composed of a rotor and a stator. The rotor can be a PM or a superconductor, and vice-versa. It this work, as a PM we have used ring NdFeB permanent magnets with maximum magnetic product around 48 MGOe and a remanence about 1.3 T and different dimensions. As high-temperature superconductors, we have used YBaCuO superconductors with different shapes purchased to Can superconductors. This superconducting material present a transition temperature about 93 K [42] so they can be cooled down using relatively inexpensive $LN_2$, which boiling's temperature is about 77 K at ambient pressure.

Two SMBs have been tested under static conditions (therefore no eddy currents appear) and presented in this paper. The first one is a SMB composed of a set of hexagon superconducting pieces as the stator and a ring PM as the rotor (see Fig. 3). Separately, we have analyzed the performance of a SMBs composed of a superconducting disks 56 mm in diameter and 15 mm in height (see Fig. 8). The rotor part is composed of two concentric rings axially magnetized with a radial gap of 2.5 mm. To hold the PMs in position, an aluminum plate (low magnetic interaction) has been manufactured. For this second SMB different configurations of the magnets are analyzed to determine the stiffer one.

In order to characterize the axial and radial stiffness of the SMBs, a dedicated test bench has been designed and built. The bench, shown in Fig. 1, is mainly composed of a metallic structure made of aluminum (1) which fixes the PM in position. Due to the magnetic properties of the aluminum alloy used, low magnetic interaction with the PM and the SC is assured. In addition, the superconductors are placed and fixed inside a $LN_2$ vessel (2) and the last assembled onto a motorized lab-jack from Thorlabs, model L490MZ (3). This lab-jack stand is intended to modify the gap (Z axis) between the PM and the SC with extreme accuracy (about 5µm) within a stroke about 55 mm. The lab-jack stand is assembled onto a linear slider (4) from Igus able to modify the radial distance between the centers of the SC and the PM (X axis). This slider is controlled in a close-loop using the position signal from a laser triangulator from Microepsilon model ILD 1402-50. In summary, overall repeatability in the radial and axial position has been estimated to be better than 100 µm.

The radial and axial forces between the PMs and the SCs have been measured by using two load cells (5) from SENEL, model SX-1, C3 precision category. They were installed at either side of the T bar that supports the PMs as shown in Fig. 1. Finally, all the electronic systems are connected to a PC and data acquisition is synchronized using Labview.

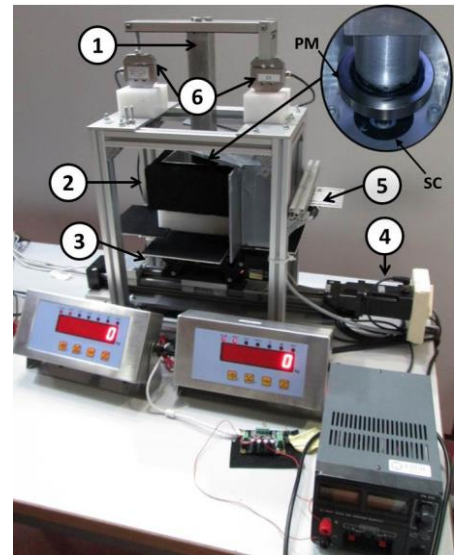

**Figure 1** - Test bench. 1) T bar, 2) $LN_2$ vessel, 3) lab-jack stand, 4) linear slider 5) laser triangulator and 6) load cells.

The first step in the measurement procedure is to define the relative axial and radial position between the PMs and the SCs. From now on, the initial axial distance (Z axis) between the SCs top surface and the PMs bottom surface will be named height of levitation or height of field cooling (HFC). Initially, both axis of the PMs and the SCs are aligned so that the radial distance (X axis) is null.

After this positioning of the parts of the SMB, both, the superconductors (contained in the $LN_2$ vessel) and the PMs, are cooled down by pouring liquid nitrogen at ambient pressure into the $LN_2$ vessel (2) shown in Fig. 1. It is well know that the magnetization of NdFeB permanent magnets is very dependent on temperature [11],[43]. By submerging the PMs too, we assure a constant temperature of the whole bearing equal to about 77 K.

Once the temperature is stable, the axial and radial position of the SC with respect to the PM is modified using the lab-jack or the slider. Drag forces will appear between the PM and the superconductors at the mixed state [44]. After each different static position (X, Z) is reached, these forces have been measured using the load cells in Fig. 1 and the axial and radial stiffness calculated from those data obtained.

A diagram of the experimental set-up is provided in Fig. 2.

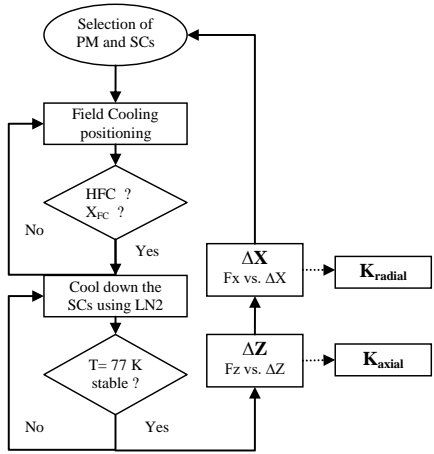

**Figure 2 -** Experimental procedure

## 4. Experimental results

In this section, the results of the experimental tests described in section 3 are presented. The first subsection 4.1 presents the results of the first kind of SMB. Subsections 4.2 and 4.3 describe the mechanical behaviour of the second kind of SMB analyzed.

*4.1 Stiffness and load capability vs. HFC*

Radial and axial stiffness have been investigated for a SMB made of 6 hexagon superconductors and a ring PM as shown in Fig. 3.

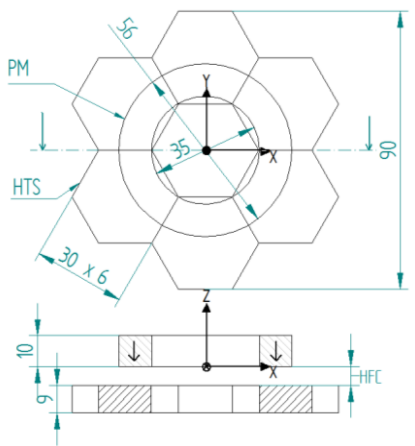

**Figure 3 -** SMB analyzed in section 4.1.

We have analyzed the levitation force provided by the bearing for different heights of field cooling (HFC) and the results are plotted in Fig. 4. This plot shows the levitation force vs. the axial displacement (Z axis) for different values of the HFC. Z position has been smoothly varied from HFC until a gap of 1.5 mm between the PM and the SC is reached. No forward motion is represented. Note that Fz vs. Z fits an exponential curve with $R^2>0.99$ in all cases.

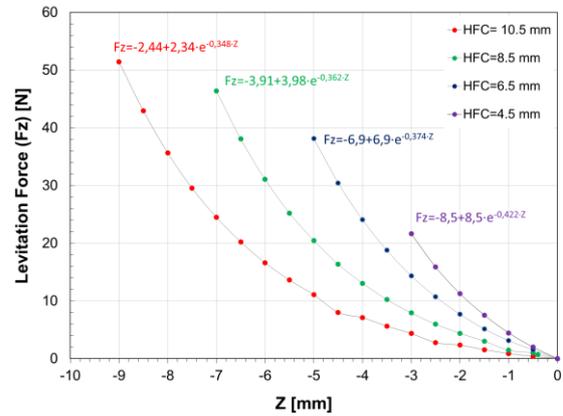

**Figure 4 -** Load capability of the SMB in Fig. 3 vs. HFC.

In order to compare the results, we have defined the load capability as the maximum levitation force provided by the bearing when the gap between the PM and the SC is equal to 1.5 mm. The load capability of the analyzed SMB is clearly enhanced when the value of the HFC is increased between 4.5 and 10.5 mm, as demonstrated in the following figure.

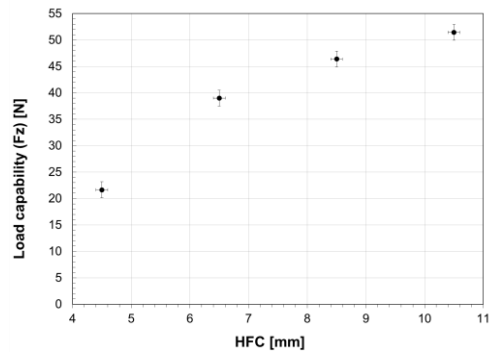

**Figure 5 -** Load capability of the SMB in Fig. 3 vs. HFC.

Similarly, we have analyzed the radial force (Fx) vs. the radial position (X axis) for different HFC. In this case, forward and backward motions are considered. The radial stiffness is clearly increased for lower values of HFC. In addition, good linearity of the radial force vs. the radial displacement is observed as shown in Fig. 6.

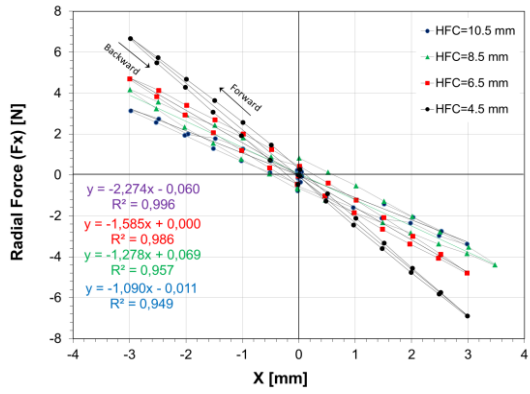

**Figure 6** - Radial force (Fx) vs. radial position of the PM (X axis) for different values of HFC. Z=-(HFC-1.5) mm in all cases.

The normalized radial and the axial stiffness and the normalized load capability for different values of the HFC are plotted in Fig. 7. For estimation of the radial stiffness, a linear regression of data has been used. In the case of the represented axial stiffness, the values of the stiffness at Z=HFC=0 mm obtained by differentiation of equations embedded in Fig. 4 have been used.

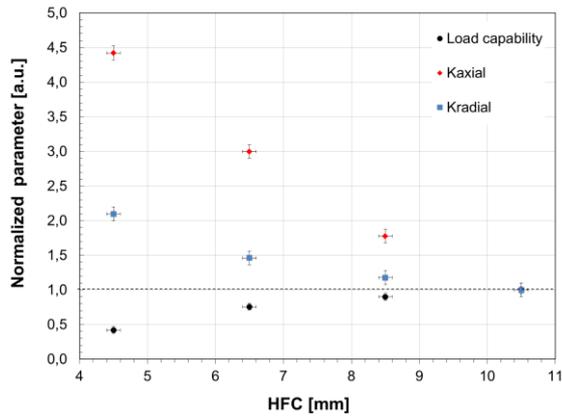

**Figure 7** - Axial, radial stiffness and load capability vs. HFC of the hexagon-shaped bearing in Fig. 3. Values are normalized at HFC= 10.5 mm in all cases.

Axial and radial stiffness seem to be exponentially decreasing functions of the HFC. However, the load capability demonstrates an exponentially increase for higher values of HFC within the margin of the situations studied.

*4.2 Influence of the polarity of the PMs in a concentric thrust bearing*

The use of multiple-pole configurations in SMB has been already proposed for the improvement of SMBs [45], [46] especially for journal bearings [10]. In this section, we propose and analyze a SMB with two concentric ring PMs. The axial and radial stiffness of the two SMBs, represented in Fig. 8, are investigated.

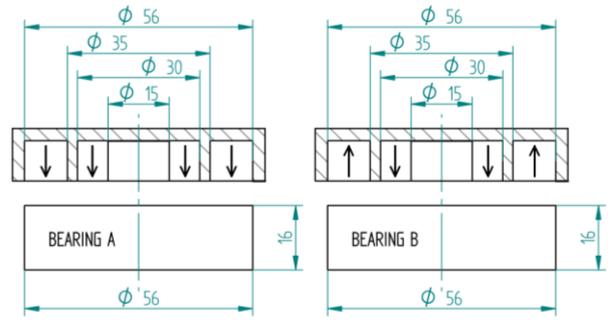

**Figure 8** - Concentric PM superconducting bearing.

Whereas in bearing A (see Fig. 8) both magnets have the same magnetization, in bearing B they have opposite ones. Relevant differences in the field distribution and intensity are expected between the two bearings. Providing high magnetic flux density and high magnetic gradients has been demonstrated to be the key to obtain larger forces and enhanced stiffness [10].

The magnetic flux density in the SC in the normal state is plotted in Fig. 9. Only half of the section is plotted due to revolution symmetry of the SMB.

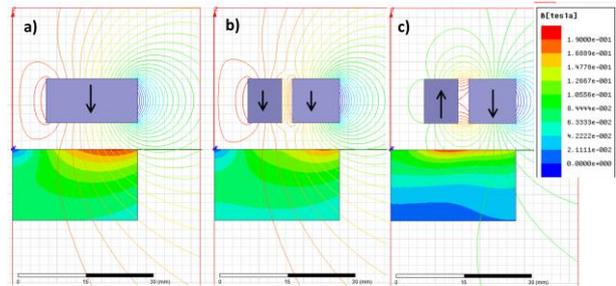

**Figure 9** - Magnetic flux density inside the SC in the normal state ($\mu_r$=1). a) PM 56 mm in outer diameter and 15 mm in inner diameter; b) bearing A, b) bearing B. Z=HFC= 6 mm in all cases.

Results of axial and radial force measurements for bearing A and bearing B are plotted in Fig. 10 and Fig. 11 respectively.

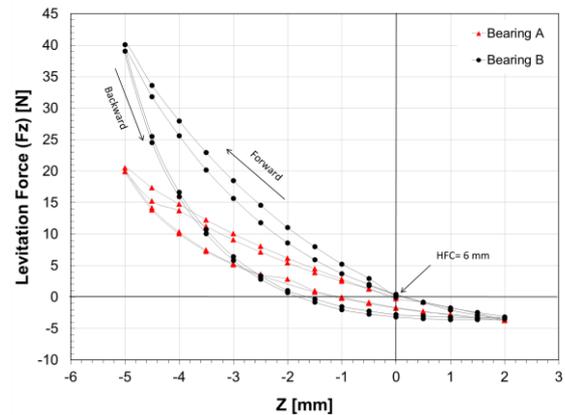

**Figure 10** - Levitation force vs. Z position for bearings A and B. HFC= 6 mm in both cases.

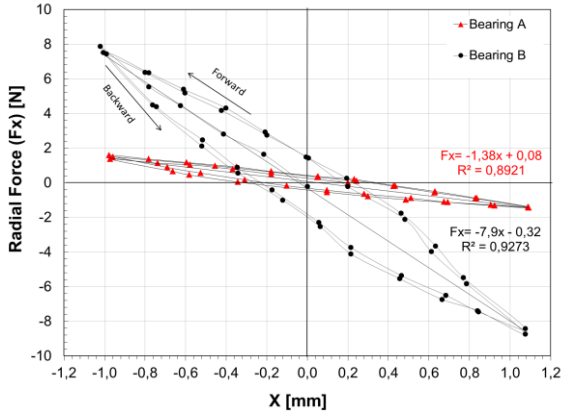

**Figure 11** - Radial force vs. X position for bearings A and B. HFC= 6 mm and Z=-5 in both cases.

In summary, bearing B shows a better overall performance in terms of stiffness and load capability. Load capability has been increased about 100% with respect to bearing A and radial and axial stiffness has been improved from about -2.5 to -4.2 N/mm and from about -1.4 to -7.9 N/mm respectively. However, it has to be considered that hysteresis is enhanced in that bearing as well.

In addition, we have observed and exponential relationship between the measured radial stiffness and the Z position of the PM. As shown in Fig. 12, the lower the values of Z, the higher the radial stiffness.

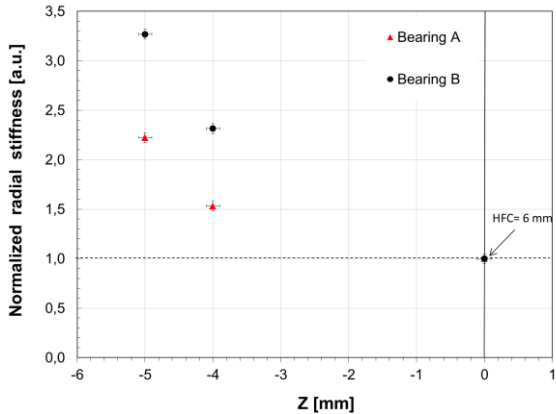

**Figure 12** - Normalized radial stiffness vs. Z position of the PM. HFC= 6 mm in all cases. Bearing A: K(z=0)=0.62 N/mm; bearing B: Kx(z=0 mm)= 2.42 N/mm. Stiffness obtained from linearization of the experimental results.

*4.3 Hysteresis and force relaxation*

Superconducting magnetic bearings at the mixed state present hysteresis. This hysteretic behavior is thought to be mainly related to changes in the magnetization of the SC when they are at the mixed state and the presence of defects in the crystalline structures and chemical impurities in the superconductors that prevent the free mobility of the magnetic vortex in the superconductor in the mixed state [47]. The hysteresis in the levitation force can be observed for example in Fig. 10 and the hysteresis in the radial force in Fig. 11.

In addition, a time-dependent relaxation of the radial and axial force can be present in a superconductor in the mixed state. As an example, normalized levitation force vs. time is shown in Fig. 13. In this experiment, the position of the PM has been modified from $Z_0=0$ to $Z=-5$ mm and then, the levitation force measured in time.

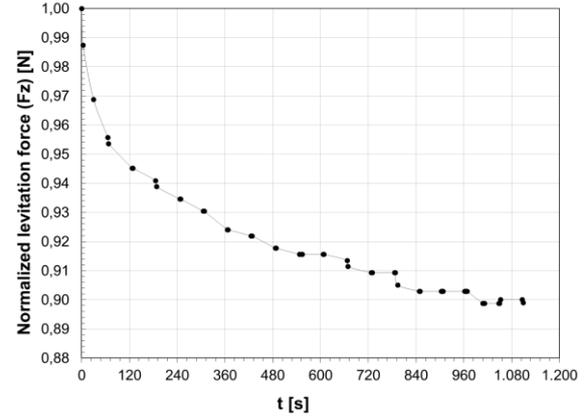

**Figure 13** - Relaxation of the levitation force in time for $\Delta Z= -5$ mm, $Z_0=0$ mm, $X= 0$ and HFC= 6 mm.

This force relaxation must be always considered for an adequate mechanical design.

## 5. Conclusions

Contactless superconducting magnetic bearings find application in fields where the lack of contact is a requirement or an advantage like for example solving all the tribological problems in cryogenic environments. However, their low load capability and stiffness is still a limitation to the spread of this technology. In this paper, a set of rules for the design of high-stiffness thrust superconducting magnetic bearings are proposed and experimentally demonstrated:

- Both, the radial and the axial stiffness are exponentially enhanced when the height of field cooling is reduced.
- However, the load capability is exponentially reduced when the height of field cooling is lowered.
- The radial stiffness in the bearing is exponentially dependent on the airgap between the PM and the SC for a given HFC.
- Multipole configurations in a superconducting bearing can drastically improve their performance. An opposite orientation of the magnetization directions highly enhances the axial and radial stiffness and the load capability of a SMB with respect to a bearing with equal orientation of the magnetization of the PMs. However, the hysteresis is enhanced as well.

**Acknowledgments**

The research leading to these results received funding from the European Community's Seventh Framework Programme (FP7/2007-2013) under grant agreement no. 263014.